\documentclass[twocolumn,showpacs,preprintnumbers,amsmath,amssymb]{revtex4}
\usepackage{graphicx}% Include figure files
\usepackage{dcolumn}% Align table columns on decimal point
\usepackage{bm}% bold math

\begin{document}

\preprint{APS/123-QED}

\title{CDF Wjj anomaly as a non-perturbative effect of the electro-weak interaction}% Force line breaks with \\

\author{Boris A. Arbuzov}

\affiliation{Skobeltsyn Institute for Nuclear Physics of Moscow State University\\ Leninskie gory 1, 119991 Moscow, Russia}%
\email{arbuzov@theory.sinp.msu.ru}
\author{Ivan V. Zaitsev}

\affiliation{Skobeltsyn Institute for Nuclear Physics of Moscow State University\\ Leninskie gory 1, 119991 Moscow, Russia}
\date{\today}% It is always \today, today,
\begin{abstract}
The recently reported CDF excess at $120\,-\, 160\,GeV$ in  invariant mass distribution of jet pairs accompanying $W$-boson~\cite{CDF} is tentatively interpreted as a bound state of two $W$ decaying to quark-anti-quark pair.  Non-perturbative effects of EW interaction obtained by application of Bogoliubov compensation approach lead to such bound state due to existence of anomalous three-boson gauge-invariant effective interaction. The application of this scheme gives satisfactory agreement with existing data without any adjusting parameter but the bound state mass $145\,GeV$. 
\end{abstract}

%Use showkeys class option if keyword
\maketitle

%\section{Effective anomalous three-boson 
%interaction}

In works~\cite{BAA04, BAA06, AVZ06, BAA07, AVZ09, BAA09, AZ11},
N.N. Bogoliubov compensation principle~\cite{Bog1, Bog2} 
was applied to studies of a spontaneous generation of effective non-local interactions in renormalizable gauge theories.

In particular, papers~\cite{BAA09,AZ11}  deal with an application of the approach to the electro-weak interaction and a possibility of spontaneous generation of effective anomalous three-boson interaction of the form
\begin{eqnarray}
& &-\,\frac{G}{3!}\cdot\,\epsilon_{abc}\,W_{\mu\nu}^a\,W_{\nu\rho}^b\,W_{\rho\mu}^c\,;
\label{FFF}\\
& &W^3_{\mu \nu}\,=\,\cos\theta_W\,Z_{\mu \nu}\,+\,\sin\theta_W\,A_{\mu \nu}\,;\nonumber
\end{eqnarray}
with uniquely defined form-factor $F(p_i)$. It was done of course in the framework of an approximate scheme, which accuracy was estimated to be $\simeq (10 - 15)\%$~\cite{BAA04}.
Effective interaction~(\ref{FFF}) is
usually called anomalous three-boson interaction and it is considered for long time on phenomenological grounds~\cite{Hag1, Hag2}. Note, that the first attempt to obtain the anomalous three-boson interaction in the framework of Bogoliubov approach was done in work~\cite{Arb92}. Our interaction constant $G$ is connected with
conventional definitions in the following way
\begin{equation}
G\,=\,-\,\frac{g\,\lambda}{M_W^2}\,;\label{Glam}
\end{equation}
where $g \simeq 0.65$ is the electro-weak coupling.
The current limitations for parameter $\lambda$ read~\cite{EW}
\begin{eqnarray}
& &\lambda\,=\,-\,0.016^{+0.021}_{-0.023}\,;\nonumber\\
& & -\,0.059< \lambda < 0.026\,
(95\%\,C.L.)\,.
\label{lambda1}
\end{eqnarray}
Interaction~(\ref{FFF}) increases with increasing momenta $p$ and 
corresponds to effective dimensionless coupling being of the following order of magnitude
\begin{equation}
g_{eff}\,=\,\frac{g\,\lambda\,p^2}{M_W^2}\,.\label{geff}
\end{equation}
Thus for sufficiently large momentum interaction~(\ref{FFF})  becomes strong 
and may lead to physical consequences analogous to that of the usual strong interaction (QCD). In particular bound states and resonances constituting of $W$-s (W-hadrons) may appear. Let us estimate the typical scale for the effect. We know that in QCD upper bound of a region of really strong interaction (non-perturbative region) is around $600\,MeV$ where $\alpha_s \simeq 0.5$ that is coupling $g_s = \sqrt{4 \pi \alpha_s} =
2.5$. So we have to equate $g_{eff}$~(\ref{geff}) to this value and define the typical value $p_{typ}$ that gives
\begin{equation}
p_{typ}\,=\,M_W\sqrt{\frac{g_{eff}}{g\,\lambda}}\,\simeq 650\,GeV\,;\label{ptyp}
\end{equation}
where we have taken for modulus of $\lambda$ its maximal value from limitation~(\ref{lambda1}).
Now we have the lightest hadron -- the pion with mass $\simeq 140\, MeV$ for typical scale $600\,MeV$ in QCD and for estimated 
$p_{typ}$~(\ref{ptyp}) we have possible mass of the lightest W-hadron
\begin{equation}
M_{min}\,=\,\frac{p_{typ}\,M_\pi}{600\,MeV}\,\simeq 150\,GeV\,;\label{min}
\end{equation}
The excess detected in work~\cite{CDF} is situated just in this region. So one might try to consider interpretation of effect~\cite{CDF} as a manifestation of a $W$-hadron. 

In the present work we apply these considerations along with some results of work~\cite{AZ11} to data indicating on an excess in jet pair production accompanied by $W$ at TEVATRON~\cite{CDF} in region of $jj$ invariant mass $120\,-\,160\,GeV$. Some indications for state with the same mass at LHC are discussed in work~\cite{143}.

Let us assume that this excess is due to existence of 
bound state $X$ of two $W$. This state $X$ is assumed to have 
spin 1 and weak isotopic spin also 1. Then vertex of $XWW$ interaction has the following form
\begin{equation}
\frac{G_X}{2}\,\epsilon_{a b c}\,W_{\mu \nu}^a\,W_{\nu \rho}^b\,X_{\rho \mu}^c\,\Psi\,;
\label{XWW}
\end{equation}
where $\Psi$ is a Bethe-Salpeter wave function of the bound state. The main interactions forming the bound state are just non-perturbative interactions~(\ref{FFF}, \ref{XWW}). This means that we take into account exchange of vector boson $W$ as well as of vector bound state $X$ itself. In diagram form the corresponding Bethe-Salpeter equation is presented in Fig. 1. For small mass $M_X$ of state $X$ we expand the kernel of the equation in powers of $M_W^2$ and $M_X^2$ and obtain the following equation 
\begin{widetext}
\begin{eqnarray}
& &\Psi(x) = \frac{G^2+G_X^2}{32 \pi^2}\int_0^{Y_0}\Psi(y) y dy - \frac{G^2+G_X^2}{32 \pi^2}\biggl(\frac{1}{12 x^2}\int_0^x \Psi(y)y^3 dy-\frac{1}{6 x}\int_0^x \Psi(y) y^2 dy - 
\frac{x}{6}\int_x^{Y_0} \Psi(y) dy+\nonumber\\
& &\frac{x^2}{12}\int_x^{Y_0}\frac{\Psi(y)}{y} dy\biggr) + \frac{g\, G}{4\, \pi^2}\biggl(\int_0^{Y_0} \Psi(y) dy- \frac{3}{8 x^3}\int_0^{x}\Psi(y) y^3 dy+ \frac{7}{8 x^2}\int_0^{x}\Psi(y) y^2 dy-\frac{1}{2 x}\int_0^{x}\Psi(y) y dy+\nonumber\\
& &\frac{x}{8}\int_x^{Y_0} \frac{\Psi(y)}{y} dy-\frac{x^2}{8}\int_x^{Y_0} \frac{\Psi(y)}{y^2} dy)\biggr) - \frac{\mu\,\bar G \sqrt{2}}{\pi}\biggl(\int_0^{Y_0} \Psi(y) dy- \frac{1}{12x^2}\int_0^{x}\Psi(y) y^2 dy+\frac{1}{6 x}\int_0^{x}\Psi(y) y dy+\label{BS}\\
& & \frac{x}{6}\int_x^{Y_0} \frac{\Psi(y)}{y} dy-\frac{x^2}{12}\int_x^{Y_0} \frac{\Psi(y)}{y^2} dy\biggr)-\frac{\chi\, \bar G \sqrt{2}}{\pi}\biggl(\frac{1}{24}\int_0^{Y_0} \Psi(y) dy-\frac{1}{192 x^3}\int_0^{x}\Psi(y) y^3 dy+\nonumber\\
& & \frac{1}{64 x}\int_0^{x}\Psi(y) y dy+\frac{x}{64}\int_x^{Y_0} \frac{\Psi(y)}{y} - \frac{x^3}{192}\int_x^{Y_0}\frac{\Psi(y)}{y^3} dy
\biggr).\quad \mu = \frac{\bar G\,M_W^2}{16 \pi\,\sqrt{2}}\,;\; \chi = \frac{\bar G\,M_X^2}{ 16 \pi\,\sqrt{2}}\,;\;\bar G=\sqrt{G^2+G_X^2}\,.\nonumber 
\end{eqnarray}
\end{widetext}
Here $x=p^2$ is the external momentum squared and $y$ is the integration momentum squared. Gauge electro-weak coupling $g$ enters due to diagrams of the second line of Fig. 1. Upper limit $Y_0$ is introduced for the sake of generality due the experience of works~\cite{BAA04, BAA06, AVZ06, BAA07, AVZ09, BAA09, AZ11}, according to which $Y_0$ may be either $\infty$ or some finite quantity. That is $Y_0$ is defined in a process of solving an equation. From the physical point of view
an effective cut-off $Y_0$ bounds a "low-momentum" region where our non-perturbative effects act and we consider the equation at interval $[0,\, Y_0]$ under condition
\begin{equation}
\Psi(Y)\,=\,0\,. \label{Y0}
\end{equation}
For interaction~(\ref{FFF}) $Y_0$ is defined in work~\cite{AZ11}. 

We shall solve equation~(\ref{BS}) by iterations. Let us perform the following substitution
\begin{equation}
z\,=\,\frac{(G^2+G_X^2) x^2}{512 \pi^2}\,,\quad t\,=\,\frac{(G^2+G_X^2) y^2}{512 \pi^2}\,; \label{zt}
\end{equation}\\
then equation~(\ref{BS}) takes the following form
\begin{widetext}
\begin{eqnarray}
& &\Psi(x)\equiv \Psi_0(z) = 8 \int_0^{z'_0}\Psi_0(t) dt - \frac{2}{3 z}\int_0^z \Psi_0(t) t dt+\frac{4}{3 \sqrt{z}}\int_0^z \Psi_0(t)\sqrt{t}\,dt +\frac{4 \sqrt{z}}{3}\int_z^{z'_0} \frac{\Psi_0(t)}{\sqrt{t}} dt -\nonumber\\ 
& &\frac{2 z}{3}\int_z^{z'_0}\frac{\Psi_0(y)}{y} dy +\frac{g'\sqrt{2}}{4\, \pi}\biggl(8 \int_0^{z'_0} \frac{\Psi_0(t)}{\sqrt{t}}\,dt-\frac{3}{z\sqrt{z}}\int_0^z \Psi_0(t) t dt +\frac{7}{z}\int_0^z \Psi_0(t)\sqrt{t} dt -\frac{4 }{\sqrt{z}}\int_0^z \Psi_0(t) dt+\nonumber\\ 
& &\sqrt{z}\int_z^{z'_0} \frac{\Psi_0(t)}{t} dt-z\int_z^{z'_0} \frac{\Psi_0(t)}{t\sqrt{t}} dt \biggr)- \mu \biggl(16 \int_0^{z'_0} \frac{\Psi_0(t)}{\sqrt{t}}\,dt-\frac{4}{3 z}\int_0^z \Psi_0(t)\sqrt{t}\, dt +\frac{8}{3\sqrt{z}}\int_0^z \Psi_0(t) dt +\label{BSZ}\\
& &\frac{8 \sqrt{z}}{3}\int_z^{z'_0} \frac{\Psi_0(t)}{t} dt-\frac{4 z}{3}\int_z^{z'_0} \frac{\Psi_0(t)}{t\sqrt{t}} dt\biggr)- \chi\biggl(\frac{2}{3} \int_0^{z'_0} \frac{\Psi_0(t)}{\sqrt{t}}\,dt-\frac{1}{12 z \sqrt{z}}\int_0^z \Psi_0(t)t dt +\frac{1}{4 \sqrt{z}}\int_0^z \Psi_0(t) dt +\nonumber\\
& & \frac{\sqrt{z}}{4 }\int_z^{z'_0} \Psi_0(t) dt-\frac{z \sqrt{z}}{12 }\int_z^{z'_0} \frac{\Psi_0(t)}{t^2} dt\biggr);\quad \Psi_0(0)\,=\,1\,;\nonumber\\
& & R\,=\,\frac{G_X^2}{G^2}\,;\quad 
z'_0 = \frac{(G^2+G_X^2) Y_0^2}{512 \pi^2}\,=\,(1+R)\,z_0\,;\quad g'=\frac{g}{\sqrt{1+R}}\,.\label{cond}
\end{eqnarray}
\end{widetext}
Here we shall use results of work~\cite{AZ11} according to which consideration of the compensation equation for effective interaction~(\ref{FFF}) we obtain definite value for $z_0 = 9.6175$ and value for $g(M_W)$ being consistent with experimental value $g(M_W) = 0.65$. Thus, we use these values in the forthcoming calculations.
\begin{equation}
z_0\,=\,9.6175\,;\quad g = g(M_W)=0.65\,.\label{z0g}
\end{equation}  

Let us formulate the first approximation to equation~(\ref{BSZ}). The first five terms of rhs of~(\ref{BSZ}) will present the simplest zero approximation. Then we substitute into terms being proportional to $g',\,\mu',\,\chi'$ a solution of the zero approximation and take into account constant terms and terms being proportional to $\sqrt{z},\,\sqrt{z}\,ln\,z$. Terms being $O(z\,ln z)$ and smaller (for small z) are neglected. Just this approximation was used in works~\cite{AZ11, BAA09}. 

The zero approximation evidently reads
\begin{equation}
\Psi_{00}(z)=\frac{\pi}{2}\,G^{21}_{15}\bigl(z|^0_{1,0,1/2,-1/2,-1}\bigr)\,.\label{Psi0}
\end{equation}
Here the solution is expressed in terms of a Meijer function~\cite{BE}.
Then set of equations~(\ref{BSZ}) takes the form
\begin{widetext}
\begin{eqnarray}
& &\Psi_0(z) = INH - \frac{2}{3 z}\int_0^z \Psi_0(t) t dt+\frac{4}{3 \sqrt{z}}\int_0^z \Psi_0(t)\sqrt{t}\,dt +\frac{4 \sqrt{z}}{3}\int_z^{z'_0} \frac{\Psi_0(t)}{\sqrt{t}} dt -\frac{2 z}{3}\int_z^{z'_0}\frac{\Psi_0(y)}{y} dy\,;\nonumber\\ 
& & INH\,=\,1-\sqrt{z}\,\biggl(\frac{g'\sqrt{2}}{8\, \pi}\,+ \frac{8\, \mu}{3}\,-\,\frac{\chi}{4}\biggr)\biggl(ln\,z + 4\,\gamma +4\,ln\,2 + \frac{\pi}{2}\,G^{21}_{15}\bigl(z'_0|^0_{0,0,1/2,-1/2,-1}\bigr)\biggr)+\label{BSln}\\
& & \sqrt{z}\,\biggl(\frac{g'\sqrt{2}}{48\, \pi}\,+ \frac{68\, \mu}{9}\,-\,\frac{25\,\chi}{32} \biggr);\nonumber\\
& &1\,=\,8 \int_0^{z'_0}\Psi_0(t) dt\,-\,\biggl(\frac{g'\,2\,\sqrt{2}}{\pi}-16\,\mu+\frac{2\,\chi}{3}\biggr) \int_0^{z'_0} \frac{\Psi_{00}(t)}{\sqrt{t}}\,dt\,\nonumber
\end{eqnarray}
\end{widetext}
where $\gamma$ is the Euler constant.  
Now we look for solutions of set~(\ref{BSln}) bearing in mind conditions~(\ref{cond}) and values~(\ref{z0g}). We have relation   
\begin{equation}
M_X\,=\,M_W\,\sqrt{\frac{\chi}{\mu}}\,;\quad M_W\,=\,80.4\,GeV\,. \label{MX}
\end{equation}
We look for the exact solution of set of equations~(\ref{BSln}) in the following form
\begin{eqnarray}
& &\Psi_0(z)=\frac{\pi}{2}\,G^{21}_{15}\bigl(z|^{0}_{1,0,1/2,-1/2,-1}\bigr)\,+\nonumber \\
& &\,C_1\,G^{21}_{15}\bigl(z|^{1/2}_{1/2,1/2,1,-1/2,-1}\bigr)\,+\nonumber \\
& &C_2\,G^{20}_{04}\bigl(z|_{1,1/2,-1/2,-1}\bigr)\,+\label{Solution} \\
& &\,C_3\,G^{10}_{04}\bigl(-z|_{1,1/2,-1/2,-1}\bigr) \,.\nonumber
\end{eqnarray}
Let us choose a solution with "experimental"$\,$ $M_X\,=\,145\,GeV$~\cite{CDF}, then we have solution with the following parameters
\begin{eqnarray}
& &C_1=-\,0.015282\,;\quad C_2=-3.26512\,;\nonumber\\ 
& &C_3=1.27962\,10^{-11}\,;\quad g'=0.03932\,; \label{Solution1}\\
& &\chi=0.0074995\,;\quad z_0'=2627.975 \,;\nonumber\\
& &\mu\,=\,0.002305\,.\nonumber
\end{eqnarray}
Parameters~(\ref{Solution1}) defines the following physical parameters
\begin{eqnarray}
& &G\,=\,\frac{0.0099}{M_W^2}\,;\quad \lambda\,=\,-\,\frac{G\,M_W^2}{g}\,=\,-\,0.0152\,;\nonumber\\ & &M_X\,=\,145\, GeV\,;\quad  |G_X|\,=\,\frac{0.1639}{M_W^2}.\label{phys}
\end{eqnarray}
Value $\lambda$~(\ref{phys}) agrees with restrictions~(\ref{lambda1}). With value of $G$ from~(\ref{phys}) we have additional solutions for "radial excitations" $X_i$ with the following masses and coupling constants
\begin{eqnarray}
& &M_{X_1}\,=\,180.7\,GeV\,;\quad |G_{X_1}|\,=\,\frac{0.0628}{M_W^2}\,;\nonumber\\
& &M_{X_2}\,=\,205.1\,GeV\,;\quad |G_{X_2}|\,=\,\frac{0.1155}{M_W^2}\,.\label{X'}\\
& &M_{X_3}\,=\,244.2\,GeV\,;\quad |G_{X_3}|\,=\,\frac{0.1837}{M_W^2}\,.\nonumber
\end{eqnarray}  
With these masses $X_{1,2,3}$ decay into pair of $W$-s, i.e.
\begin{eqnarray} 	
& &X^{\pm}_{1,2,3}\,\to\,W^\pm\,+(Z, \gamma)\,;\quad X^0_{1,2,3}\,\to\,W^+\,+W^-\,;\nonumber\\
& &\Gamma(X_{1}^{0})\,=\,0.0086\,GeV\,;\quad \Gamma(X_{1}^{\pm})\,=\,0.0051\,GeV\,;\nonumber\\
& &\Gamma(X_{2}^0)\,=\,0.126\,GeV\,;\quad \Gamma(X_{2}^\pm)\,=\,0.083\,GeV\,;\nonumber\\
& &\Gamma(X_{3}^0)\,=\,1.26\,GeV\,;\quad \Gamma(X_{2}^\pm)\,=\,0.89\,GeV\,;\label{decX'}\\
& & BR(X^+_{1}\to W^+Z) = 0.44;\; BR(X^+_{1}\to W \gamma) = 0.56.\nonumber\\
& & BR(X^+_{2}\to W^+Z) = 0.80;\; BR(X^+_{2}\to W \gamma) = 0.20.\nonumber\\
& & BR(X^+_{3}\to W^+Z) = 0.91;\; BR(X^+_{3}\to W \gamma) = 0.09.\nonumber
\end{eqnarray}

Now interaction~(\ref{XWW}) with parameters~(\ref{phys}) defines reactions of $X^\pm,\,X^0$ production at TEVATRON and their decays.
 Bound states $X$ interact with fermion doublets $\psi_L$ due to diagram presented at Fig. 2. The effective interaction is 
described by the following expression 
\begin{eqnarray}
& &L_{X \psi}\,=\,g_X\, X^a_\nu\, \bar\psi_L\,\tau^a\,\gamma^\nu\psi_L;\label{GXU}\\
& &g_X = \frac{g^2\,G_X\,M_X^2}{64\,\pi^2} \int_{\mu^2}^{z'_0}\frac{\Psi_0(t)}{t}\,dt = 0.000879\,.\nonumber
\end{eqnarray}
Due to interactions~(\ref{XWW}, \ref{GXU}) there are the following decays of bound states $X$ (in calculations of decay parameters and cross-sections we use CompHEP package~\cite{Boos})
\begin{eqnarray}
& &X^\pm\,\to\,W^\pm\,+\,\gamma\,( 85.9\%);\quad X^\pm\,\to\,j\, j\;(9.5\%);\nonumber\\
& &X^0\,\to\,j\, j\;(71.4\%)\,;\label{decay}
\end{eqnarray} 
where we associate a jet with each quark. The states are narrow, but small total widths do not contradict to data~\cite{CDF} because the observed width of the enhancement corresponds to experimental resolution. One has also to bear in mind that real masses of neutral and charged $X$ may differ by few $GeV$.
   
For estimation of $X$ production cross-sections we have to take into account that according to EW gauge invariance isotopic triplet $X^a$ necessarily interacts with gauge field $W^a$ and the interaction vertex is just the gauge one with the same coupling $g$
\begin{eqnarray}
& &\Gamma_{\mu \nu \rho}^{abc}(p,q,k) = g\, \epsilon^{abc}\Bigl(\Phi_\kappa(p,q,k)\,\kappa(g_{\nu\rho} k_\mu-g_{\rho\mu}k_\nu)+\nonumber\\
& &\Phi_g(p,q,k)\,\bigl(g_{\mu\nu}(q_\rho-p_\rho)-g_{\nu\rho}q_\mu+g_{\rho\mu}p_\nu\bigr)\Bigr);\label{gauge}
\end{eqnarray}
where $\Phi_{\kappa, g}(p_i)$ are form-factors, which are equal to unity for $W$ momentum $k = 0$ and other two momenta $p,\,q$ on the mass shell, $\kappa$ is the well-known parameter describing quadrupole interaction of a vector particle. In the present approximation $\kappa\,=\,0$. Effective total energy for partons collisions at TEVATRON is around $300\,GeV$ which is essentially smaller than typical value~(\ref{ptyp}). Thus, for estimates of cross-sections at TEVATRON we take $\Phi_g = 1$ and $\Psi\,\simeq \Psi(0)\,=\,1$. 
   
So taking into account all relevant interactions~(\ref{XWW},\ref{GXU},\ref{gauge}) we obtain the following estimates for cross-sections for energy $\sqrt{s}=1960\,GeV$
\begin{eqnarray}
& &\sigma(p \bar p\to W^{\pm}X^{0}+...)= 1.86\,pb\,;\nonumber\\
& &\sigma(p \bar p\to W^\mp\,X^{\pm}+...)= 1.71\,pb\,;\nonumber\\
& &\sigma(p  \bar p\to Z\,X^{\pm}+...)= 1.37\,pb\,;\label{pbarpX}\\
& &\sigma(p \bar p\to X^0\,X^{\pm}+...)= 0.35\,pb\,;\nonumber\\
& &\sigma(p  \bar p\to X^\mp\,X^{\pm}+...)= 0.26\,pb\,.\nonumber
\end{eqnarray} 
Taking into account branching ratios~(\ref{decay}) we obtain for additional $Wjj$ and $Zjj$ production in the region of enhancement the following estimate. We also divide cross-section for $jet-jet$ production into two parts: with accompanying $\gamma$ and without $\gamma$
\begin{eqnarray}
& &\sigma(p \bar p\to W^{\pm} + \gamma +2\,j+...)= 0.26\,pb\,;\nonumber\\
& &\sigma(p \bar p\to W^{\pm} +2\,j+...) = 1.49\,pb\,;\label{jj}\\
& &\sigma(p \bar p\to Z+2\,j+...) = 0.13\,pb\,.\nonumber
\end{eqnarray}\\ 
Total cross-section for $Wjj+Wjj\gamma$: $\sigma(Wjj) = 1.75\,pb$ occurs to be rather smaller than result~\cite{CDF} $\sigma (Wjj) = 4.0 \pm 1.2\,pb$,  whereas  small value for $Zjj$ production quite agrees with~\cite{CDF} data. However recent results of $D0$~\cite{D0} do not support large value for $\sigma (Wjj)$ and give upper limit for cross-section under study $\sigma (Wjj) < 1.9\, pb$ (95\% C.L.). As a matter of fact our result~(\ref{jj}) evidently does not contradict both results, because it differs from CDF number~\cite{CDF} by less than two s.d..

Processes~(\ref{pbarpX}) contribute also to observable reactions of pair weak boson production. From~(\ref{decay}, \ref{pbarpX}) we have additional contribution
\begin{equation}
\Delta \bigl(\sigma(p \bar p\to W^+W^-)+\sigma(p \bar p\to Z W^\pm) \bigr)= 2.8\,pb\,;\label{WW}
\end{equation}\\ 
that gives no contradiction to data~\cite{CDF1} at TEVATRON 
\begin{eqnarray}
& &\sigma( W^+\,W^-)+\sigma( Z\,W^\pm)= 17.4 \pm 3.3\,pb\,;\label{WW1}\\
& &(\sigma( W^+\,W^-)+\sigma( Z\,W^\pm))_{SM}  = 15.1 \pm 0.9\,pb\,;\nonumber
\end{eqnarray} 
where the lower line corresponds to SM NLO calculations~\cite{NLO}. 
We see that possible contribution~(\ref{WW}) comfortably fits into error bars of difference between experimental and theoretical (SM) numbers~(\ref{WW1}).

The production of radial excitations $X_i$ may be compared with data on search of resonant $WW$ and $WZ$ production~\cite{WZ}. The 
results following from values of parameters~(\ref{X'}, \ref{decX'}) are the following ($B_i = BR(X_i\to W\,Z)$)
\begin{eqnarray}
& &X_1: \sigma B_1= 0.15\,pb\,;\;X_2: \sigma B_2= 0.76\,pb\,;\nonumber\\
& &X_3: \sigma B_3= 1.64\,pb\,.\label{SB}
\end{eqnarray}
These results by no means contradict upper limits of work~\cite{WZ}. Note that $X_i$ production is accompanied by 
additional boson either $W$ or $Z$. Thus we predict 
effects in triple weak boson production: $W^\pm\,W^+\,W^-,\,W^+\,W^-\,Z,\, W^\pm\,Z\,Z$, which are connected with $X_i$ production.  

Process $p+p\to W^\pm+\gamma+...$ was studied at LHC for  
energy $\sqrt{s}=7\,TeV$~\cite{Wgamma}. The results in comparison to SM calculations are the following
 \begin{eqnarray}
& &\sigma( W^\pm \gamma)= 56.3 \pm 5.0(st)\pm 5.0(sy)\pm 2.3(lu)\,pb\,;\nonumber\\
& &\sigma( W^\pm \gamma)_{SM}  = 49.4 \pm 3.8\,pb\,.\label{Wgamma}
\end{eqnarray}
The cross-sections for  production of $X^\pm$ and $X^0$ at LHC are estimated to be
 \begin{eqnarray}
& &\sigma(p\,p\,\to W^\pm X^0+...)\simeq 7.9 \,pb\,;\nonumber\\
& & \sigma(p\,p\,\to W^\pm X^\mp+...)\simeq 4.7 \,pb\,;\nonumber\\
& &\sigma(p\,p\,\to Z X^\pm+...)\simeq 5.7 \,pb\,;\label{ppX}\\
& & \sigma(p\,p\,\to X^\mp+...)\simeq 0.8 \,pb\,;\nonumber\\
& &\sigma(p\,p\,\to  X^0...)\simeq 0.6 \,pb\,.\nonumber
\end{eqnarray}
These results are just estimation by an order of magnitude due to significant influence of form-factors in interactions~(\ref{XWW}, \ref{gauge}) at energy of LHC. In calculations we have used average values of form-factors in the region corresponding to  the most probable $s_{eff}$ of partons: $\sqrt{s_{eff}}\,\simeq\,700\,GeV$~\cite{CTEQ}.
Additional contribution from processes~(\ref{ppX})  to $W^\pm\,\gamma$ production reads
\begin{equation}
\Delta \sigma( W^\pm \gamma)\simeq
 9.7\,pb\,.\label{Wgammaour}
\end{equation}
We see that here we also have no contradiction with data~(\ref{Wgamma}). Let us emphasize that this process is quite promising for checking of our scheme, because we not only predict additional contribution~(\ref{Wgammaour}) but we insist that this additional contribution means production of narrow resonance $X^\pm$ with mass around $145\,GeV$ which decays mostly to $W^\pm +\gamma$.

Let us present also estimate for production of the $145\,GeV$ excess at LHC for $\sqrt{s}\,=\,7\,TeV$, which follows from processes~(\ref{ppX})       
\begin{eqnarray}
& &\sigma(p\,p\,\to\,j\,j\,+ ...)\simeq
 0.5\,pb\,;\label{X0}\\
& &\sigma(p\,p\,\to\,j\,j\,+\,W^\pm\,+ ...)\simeq
 6.1\,pb\,.\nonumber
\end{eqnarray}

To conclude we once more would emphasize that the would-be resonances being considered here might serve as an evidence for 
existence of the new family of particles with mass scale of hundreds $GeV$ -- W-hadrons.

\newpage
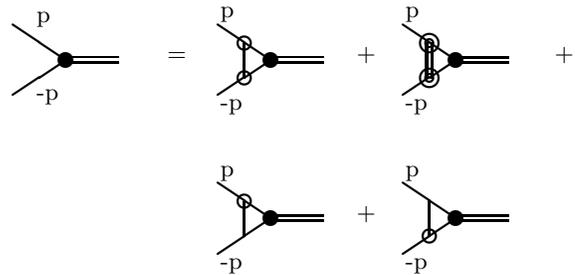
\begin{figure}
\begin{picture}(250,140)
{\thicklines
\put(35,100.5){\line(-3,2){20}}
\put(35,100.5){\line(-3,-2){20}}
\put(35,100.5){\circle*{6}}}
\put(35,101.5){\line(1,0){20}}
\put(35,99.5){\line(1,0){20}}
\put(24,114.5){p}
\put(24,85.5){-p}
%\put(44,44.5){0}
\put(73.5,100){=}
{\thicklines
\put(112.5,100.5){\line(-3,2){20}}
\put(112.5,100.5){\line(-3,-2){20}}
\put(93.5,116.5){p}
\put(93.5,81.5){-p}
%\put(122.5,44.5){0}
\put(112.5,100.5){\circle*{6}}
\put(102.5,93.5){\line(0,1){13.4}}
\put(102.5,93.8){\circle{5}}
\put(102.5,107){\circle{5}}}
%\put(2.5,80.5)
%{\line(1,-1){5}}
\put(112.5,101.5){\line(1,0){20}}
\put(112.5,99.5){\line(1,0){20}}}
\put(145,100){+}
{\thicklines
\put(182.5,100.5){\line(-3,2){20}}
\put(182.5,100.5){\line(-3,-2){20}}
\put(163.5,116.5){p}
\put(163.5,81.5){-p}
%\put(192.5,44.5){0}
\put(182.5,100.5){\circle*{6}}
\put(173.5,93.5){\line(0,1){13.4}}
\put(171.5,93.5){\line(0,1){13.4}}
\put(172.5,93.8){\circle{7}}
\put(172.5,93.8){\circle{3}}
\put(172.5,107){\circle{7}}
\put(172.5,107){\circle{3}}
%\put(2.5,80.5)
%{\line(1,-1){5}}
\put(182.5,101.5){\line(1,0){20}}
\put(182.5,99.5){\line(1,0){20}}
\put(220,100){+}
{\thicklines
\put(112.5,40.5){\line(-3,2){20}}
\put(112.5,40.5){\line(-3,-2){20}}
\put(93.5,56.5){p}
\put(93.5,21.5){-p}
%\put(122.5,44.5){0}
\put(112.5,40.5){\circle*{6}}
\put(102.5,33.5){\line(0,1){13.4}}
\put(102.5,33.8){\circle*{1}}
\put(102.5,47){\circle{5}}}
%\put(2.5,80.5)
%{\line(1,-1){5}}
\put(112.5,41.5){\line(1,0){20}}
\put(112.5,39.5){\line(1,0){20}}}
\put(145,40){+}
{\thicklines
\put(182.5,40.5){\line(-3,2){20}}
\put(182.5,40.5){\line(-3,-2){20}}
\put(163.5,56.5){p}
\put(163.5,21.5){-p}
%\put(192.5,44.5){0}
\put(182.5,40.5){\circle*{6}}
%\put(173.5,33.5){\line(0,1){13.4}}
\put(172.5,33.5){\line(0,1){13.4}}
\put(172.5,33.8){\circle{5}}
%\put(172.5,33.8){\circle{3}}
\put(172.5,47){\circle*{1}}
%\put(172.5,47){\circle{3}}
%\put(2.5,80.5)
%{\line(1,-1){5}}
\put(182.5,41.5){\line(1,0){20}}
\put(182.5,39.5){\line(1,0){20}}
\end{picture}
\caption{Diagram representation of Bethe-Salpeter 
equation for W-W bound state. Black spot corresponds to BS wave function. Empty circles correspond to point-like anomalous
three-gluon vertex~(\ref{FFF}), double circle -- XWW vertex~(\ref{XWW}). Simple point -- usual gauge triple $W$ interaction. Double line -- the bound state $X$, simple line -- W. Incoming momenta are denoted by the corresponding external lines.}
\label{fig:1}       % Give a unique label
\end{figure}
\newpage
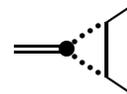
\begin{figure}
%\resizebox{0.75\textwidth}{150pt}{\includegraphics{}}
%\begin{picture}(300,25)

\begin{picture}(200,60)
{\thicklines
\put(72.5,36.5){\line(-1,0){20}}
\put(72.5,34.5){\line(-1,0){20}}
\put(87.5,45.5){\line(3,2){10}}
\put(87.5,25){\line(3,-2){10}}
\multiput(72.5,36.5)(3,2){5}{\circle*{2}}
\multiput(72.5,34.5)(3,-2){5}{\circle*{2}}
\put(87.5,25){\line(0,1){20.5}}}
%\put(105.5,42.5){p}
\put(72.5,35.5){\circle*{6}}
%\put(82.5,53.5){\line(0,1){13}}
%\put(140.5,42.5){-p}}
\end{picture}\\
 
\caption{Diagram for vertex $X\, \bar q\,q$. Dotted line -- W, double line -- bound state $X$, simple line -- a quark. Black spot -- the $XWW$ BS wave function.}
\label{fig:2}       % Give a unique label
\end{figure} 

\end{document}